# An Endangered National Heritage Site - The Cape Observatory

*IS Glass (SAAO*

## Abstract

The SAAO Cape Town campus was declared a National Heritage Site in December 2018, just short of its 200th anniversary, but is now in a run-down condition. As the former Royal Observatory, it is the oldest scientific institution in South Africa and probably in all Africa. It has a fascinating and well-documented history and surely deserves better. For many years maintenance has been neglected and many of the old telescopes and buildings are in a poor state. They are beginning to show signs of serious decay. Some examples are given.

## Introduction

Many observatories around the world with comparable histories and heritages to that of the Cape have been considering application for World Heritage Status (see https://whc.unesco.org/en/astronomy/). In fact, the Observatories of the Kazan Federal University, plausibly of less significance than ours, have this year been added to the UNESCO World Heritage List.
(see https://whc.unesco.org/en/documents/200411).

South Africa has only 10 sites on the World Heritage List and none at all of scientific importance. I believe that the Cape Observatory is a prime candidate for inscription but it needs serious refurbishment and long-term maintenance. In the following, some of the conservation issues that should be faced are highlighted, mainly those concerning the chief surviving instruments.

Apart from anything else, the National Research Foundation, the current proprietor of the site, have an obligation under Section 9 of the National Heritage Resources Act No 25 of 1999 to maintain it properly.

# The Astrographic telescope (Grubb 1890)

Recent History: The Astrographic telescope and guider were removed from the mount in 1995 and their lenses were placed in the small darkroom on the observing floor. The original Astrographic and guider telescopes are on the floor at the side of the dome. In their place on the mount is a 16-inch Parkes Newtonian that was used in the 1990s by a graduate student for infrared studies but has not been touched for over a decade.

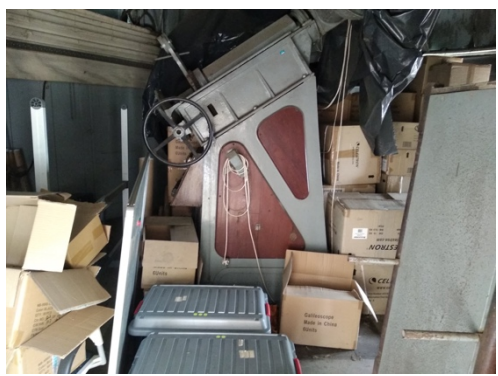 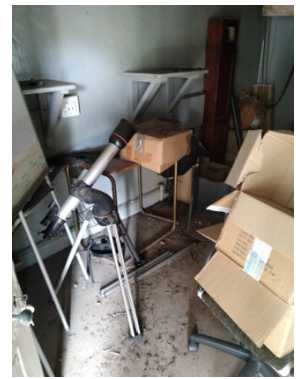

*Fig 1 (a,b): The Astrographic dome is currently used as a junk room, containing many boxes of unused equipment. These are gathering dust and dirt.*

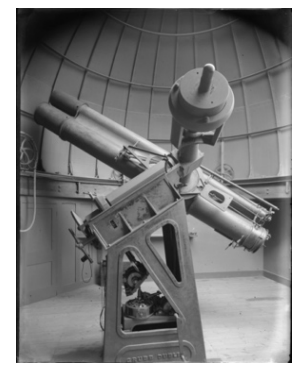

*Fig 2: The telescope in its original form. Originally constructed for photographing the Cape Zone (between declinations -40° and -52°) of the international Carte du Ciel project, it has had an interesting and varied history.*

For some details of this telescope and its achievements, please see:
http://www.saao.ac.za/~isg/poster_astrographic.pdf

Historically, among other things, the Astrographic was used for the Cape contribution to the pioneering international project "Carte du Ciel", regarded as the inspiration for the formation of the International Astronomical Union, and was used during the first detection of oxygen in stars by Frank McClean (see Gill, 1900).

The last use of the complete telescope was probably to photograph the field of SN1987A to compare with one taken with the same instrument pre-outburst. These photographs were widely circulated.

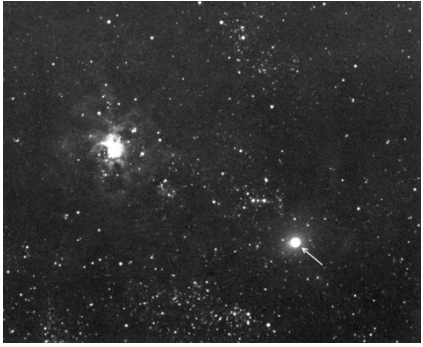 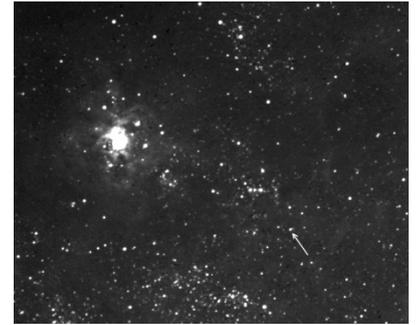

*Figs 3 and 4: Plates of the 30 Dor Region showing SN 1987A when brightest and Right: the same region showing the precursor of the supernova, Sk −69° 202. Both were taken with the Astrographic (thanks to the late Mr J. Churms).*

**The Lyot telescope (1957)**

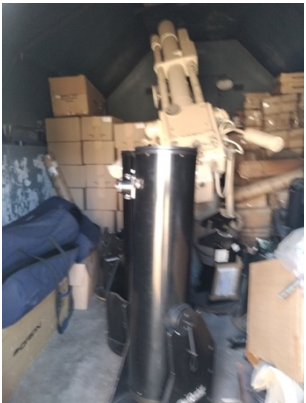

*Fig 5: The Lyot is used as storage for telescopes used in outreach and today also contains many boxes of unused material*

This is, or rather was, an automatic telescope that photographed the Sun through a H$\alpha$ filter every 2 minutes. It was installed in conjunction with the International Geophysical Year in 1957 and continued to monitor solar activity until *ca* 1980. It was constructed by Société d'Études et de Construction d'Appareillages Scientifiques et Industriels, Bordeaux, France.

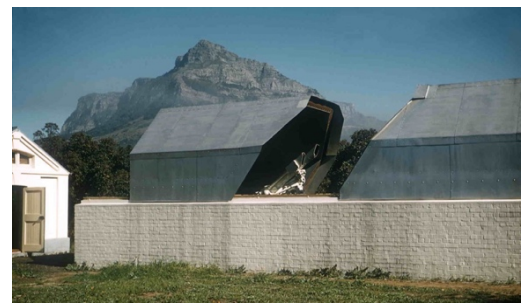

It is no longer functional. However, in terms of historical interest, it is probably less valuable than the other instruments on site.

*Fig 6: The Lyot Coronagraph In better days. This telescope followed the Sun automatically from dawn to sunset. The H$\alpha$ filter was thermostatically controlled.*

## The McClean or Victoria Telescope (Grubb, ca 1896)

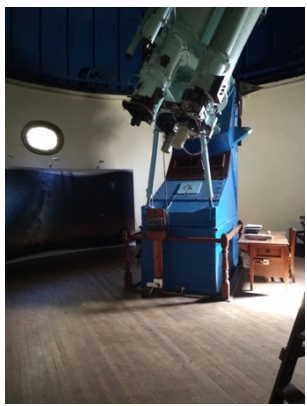

*Fig 7: This 24-inch telescope was the largest refractor in the Southern Hemisphere when constructed. It was mainly used for spectroscopy until 1926 and afterwards for stellar parallaxes until ca 1980. It remains usable.*

The building was designed by the famous architect Herbert Baker.

Its rising floor needs to be maintained periodically and such occasional attention as it has received in recent years is thanks to the efforts of Wim Filmalter, a retired engineer who is a member of the Astronomical Society of Southern Africa. He has made some recommendations on improving the performance of the hydraulic system whose seals currently tend to develop leakage problems.

This telescope is frequently used on open nights and the rising floor is something that people are usually surprised by and remember for many years afterwards. The telescope itself and its building are also showing signs of age and neglect.

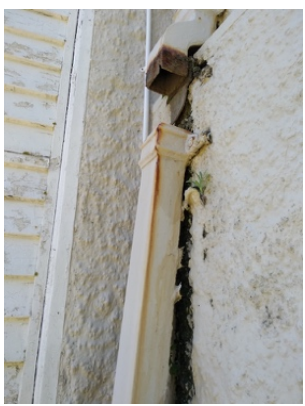 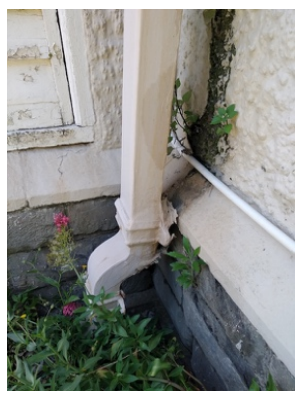 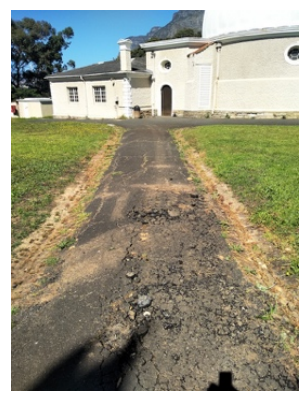

*Figs 8, 9: (left and centre) Original cast iron drain pipes becoming detached from wall. Mould growing on wall and weeks rooting in wall.*

*Fig 10: (right) path to dome broken up by recent truck traffic.*

The building also contains the Astronomical Museum of the SAAO, maintained on a voluntary basis. All items have been photographed and inventoried.

Among other problems is the messing by starlings that enter by small gaps in the shuttering.

The security of the building is a constant concern.

*Fig 11: The grills over the basement windows are loose and will not deter any determined intruder.*

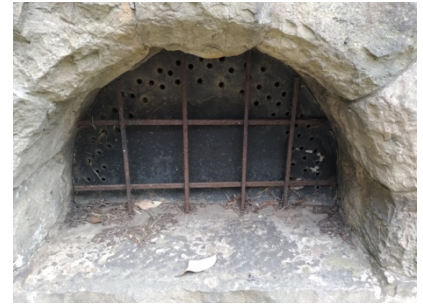

More information about the telescope and its history can be found at:
http://www.saao.ac.za/~isg/poster_mcclean.pdf

**The Reversible Transit Circle (RTC, Troughton & Simms, ca 1905)**

This instrument was one of the most refined transit circles ever built and embodies many design features that were copied in later instruments elsewhere in the world, resulting from the experience of Sir David Gill. It was of importance to the networks of positional standard stars in the "Fundamental Katalogs" of the 20th Century, only superseded by the Hipparcos and Gaia astrometric satellites of the European Space Agency

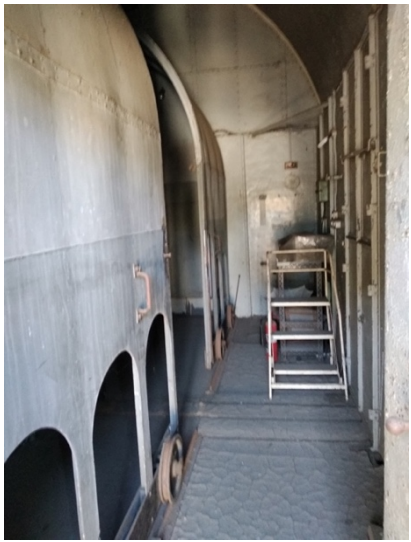 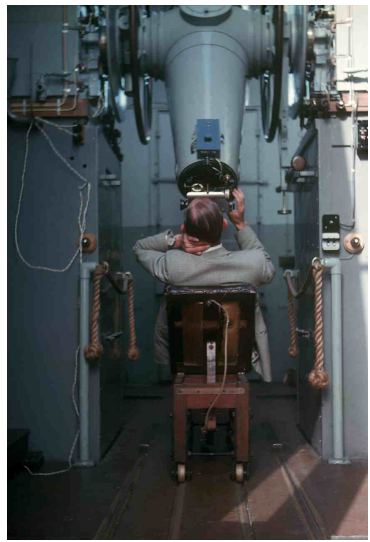 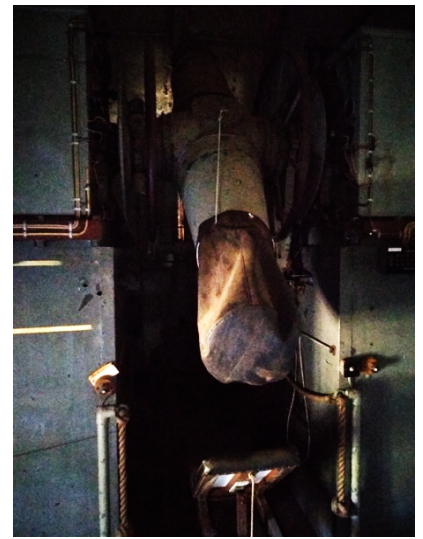

*Figs 12, 13 & 14: The interior of the RTC structure in the recent and more distant past.*

For further details, please see:
http://www.saao.ac.za/~isg/poster_gill_RTC2.pdf

The Gill circle building appears to have been stripped of auxiliary items such as clocks, thermometer and barometer, though other accessories remain. The instrument itself is not in very bad condition though probably unusable.

The building is of steel and is rusting away, particularly since **the ventilators on top are broken and admit** rain to the space between the inner and outer shell of the building. The door is jammed shut (2023) but is perhaps openable with a crowbar.

WP Koorts and I replaced the rotting wooden access steps around 2020. At this time we removed literally bucketfuls of dirt and dust.

Electricity seems no longer to be available in the building although it was functional in 2020. The dome no longer opens, having been damaged through carelessness a few years ago. [The motor starter failed during closing and a tractor was used to push the building closed. This was in ignorance of the manual override. The dome is currently jammed shut].

I believe the building could be preserved with minimal expense – the first item being to prevent the ingress of rain

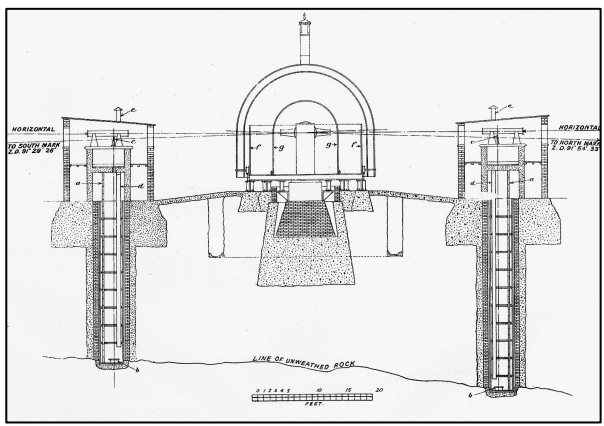

*Fig 15: The building is mostly made of sheet steel and consists of two layers. Thanks to the missing ventilator top on the West side, rain is getting in between the two layers. The remains of the missing ventilator are inside the building.*

*Fig 16 a, b & c: The damaged ventilators that let water in between the layers of the roof.*

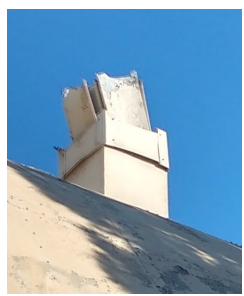 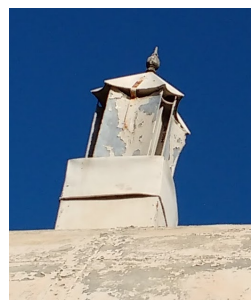 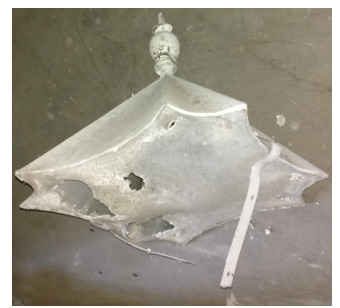

*Fig 17: Typical examples of rust on the outside of the building.*

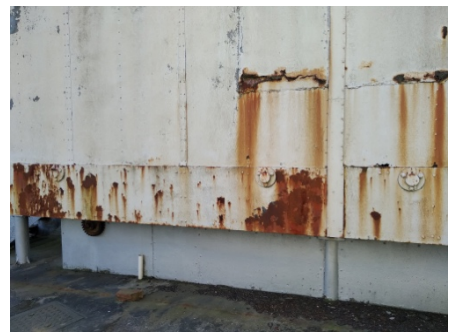

The outer buildings of the installation – the collimator houses and mark houses – are insecure and filthy, unprotected from weather, and their deep shafts (see Fig. 15) are easily accessible by children and extremely dangerous. Their keys are probably lost.

**The 18-inch telescope (various 1848-1955)**

This telescope, dating from 1855, was alost exclusively used by the late Dr AWJ Cousins and is intimately associated with the history of stellar photometry. Here the *UBVRI* Cousins system was defined.

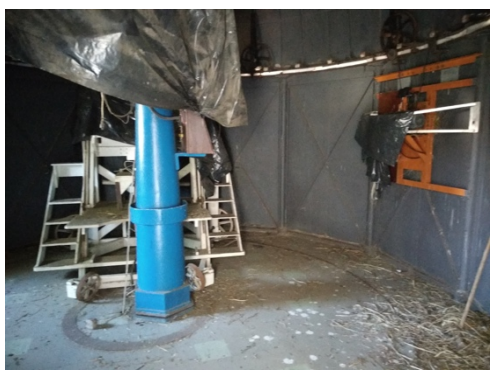

*Figs 18 & 19: The interior of the building, showing the result of bird invasion and leaks.*

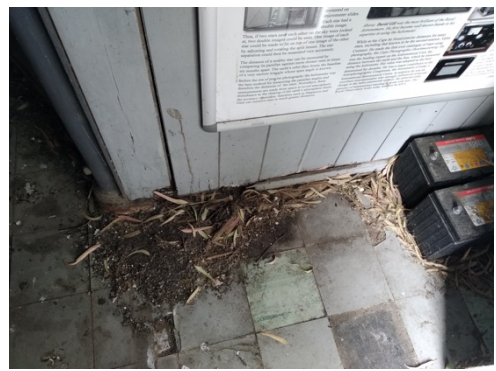

*Fig 20: The building has been invaded by starlings and the floor is covered by straw and bird droppings.*

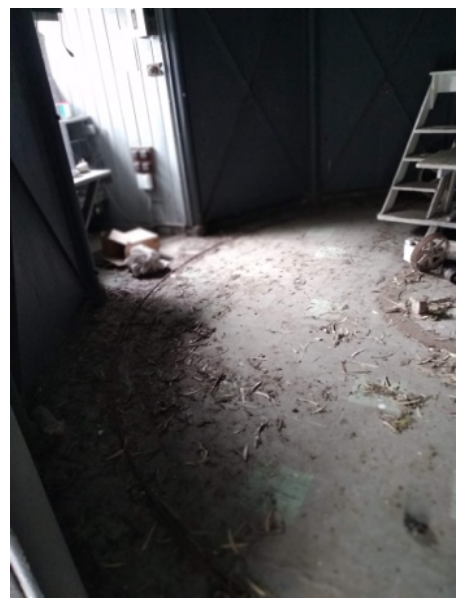

Though an attempt was made to seal it off, this was not successful. No cleaning has been performed for at least 2 years. There is a **water leak** and this also causes wetness in the archives overflow storeroom below, some of whose shelves can no longer be used. Just outside this part of the dome is a narrow downpipe whose purpose is not clear and it may be part of the problem. A large downpipe to the NW is in danger of falling off.

Some of the original photometric equipment used with the 18-inch telescope is still in the dome.

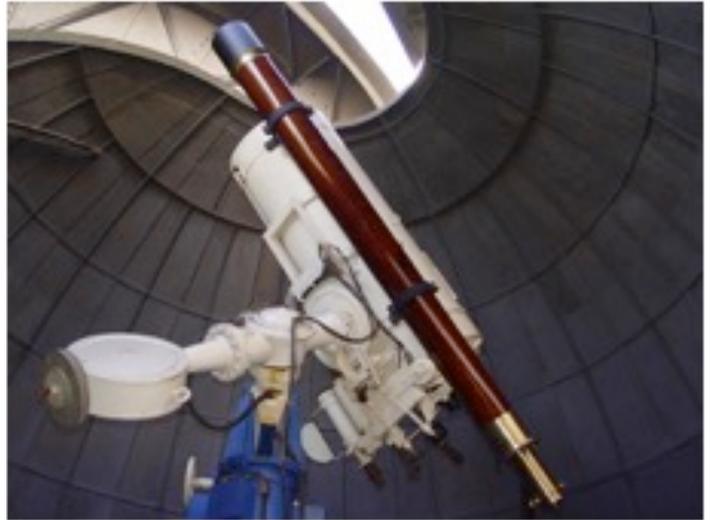

*Fig 21: In better days. The 7-inch 1848 Merz telescope is used as a guide telescope for the 18-inch reflector, seen just after it was restored.*

This building originally housed the Repsold Heliometer, dating from 1890. This was an instrument mainly used for the determination of stellar parallaxes (and consequently star distances) before the advent of photographic astrometry. The telescope mount was adapted from the Repsold one.

For the history of these telescopes and the dome, see:
 http://www.saao.ac.za/~isg/poster_18in.pdf

**The 6-inch telescope (Grubb 1882)**

This telescope was operable in recent times and may still be so.

See http://www.saao.ac.za/~isg/six-inch.jpg  for details.

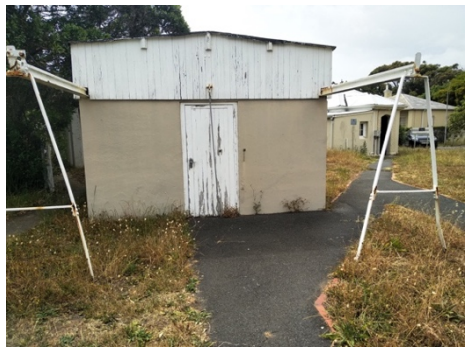 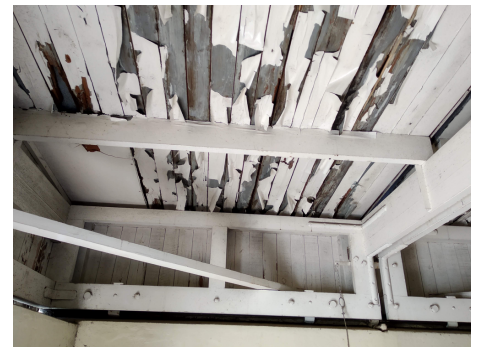

*Figs 22 & 23: It has not been cleaned recently. The woodwork of the sliding-roof building, which dates from 1935, is rotting away and has not been painted in recent years.*

**Photoheliograph (Dallmeyer 1875)**

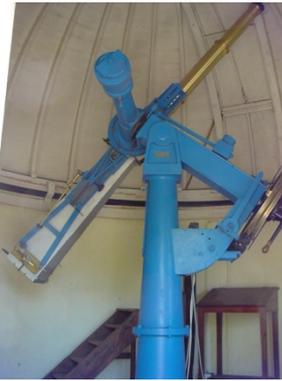

*Fig 24: This telescope is currently on a Troughton & Simms mount.*

The telescope was one of several constructed by the Dallmeyer Company of London according to the precepts of the solar pioneer Warren de la Rue. It was housed at first in a wooden building that has since been demolished. It was upgraded *ca* 1910. Two plates of the Sun were taken daily and sent to the Royal Greenwich Observatory to be used in compiling the Sunspot Index.
It is in a wooden dome of ca 1849 that runs on cannon balls.

The telescope is in working condition for solar spot viewing though not kept clean. The woodwork of the dome is in need of repair in various places. The paint is peeling. The telescope itself was cleaned and overhauled in 2009.

*Fig 25: One of the cannon ball "bearings" of the dome.*

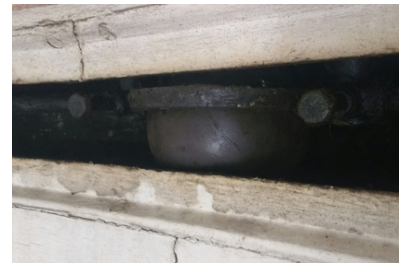

See http://www.saao.ac.za/~isg/Photoheliograph_lowres.pdf

**Some Suggestions for rehabilitation**

With the new SAAO Cape Town Visitors' Centre approaching completion, hopefully before the General Assembly of the International Astronomical Union in August 2024, now is a good time to take remedial action and restore the dignity of this special place.

While it hardly makes sense to allow unsupervised crowds access to the Observatory buildings and grounds, a lot could be done to make the Observatory a presentable National Heritage Site, interesting and pleasant to visit by those who are seriously interested. Certain places like the Museum, McClean dome and Library have long been open to the public during guided tours and the Saturday night experience, but fairly strict supervision has had to be exercised.

- Broken up pavements could be repaired, especially the old brick gutters that must have looked attractive before a century or more of general neglect, natural decay, damage by trucks, excavations for cables, weed encrustation etc.

- McClean: The rising floor is such an amazing feature that it would be well worthwhile to upgrade its hydraulics. Mr Filmalter has suggested how this could be done.

- Astrographic: The building should be cleaned up and the original telescopes replaced on the mount. The screws and fittings are probably all still there.

- 18-inch: Re-seal the dome, fix roof leak (s), re-tile floor and set up as a display of photometry.

- RTC: Repair vents, treat rust, consider making housing opening mechanism operable. Set up a display of astrometry.

- Lyot: Clean up.

- 6-inch: replace/repair roof

- 1849 dome/Photo heliograph: Repair woodwork.

- All: clean regularly. Paint as necessary.

**Reference**

Gill, Sir D., 1900. *Proc. Roy. Soc.* **65**, 413.